\documentstyle[twocolumn,floats,psfig,aps]{revtex}
\newlength{\mpclength}

\begin{document}
\title{Transverse flow of nuclear matter in
collisions of heavy nuclei at intermediate energies }
\author{Ch. Hartnack and
J. Aichelin }
\address{
SUBATECH \\
Laboratoire de Physique Subatomique et des
Technologies Associ\'ees \\
University of Nantes - IN2P3/CNRS - Ecole des Mines de Nantes \\
4 rue Alfred Kastler, F-44072 Nantes, Cedex 03, France}

\author{\begin{quote}
\begin{abstract}
{The Quantum Molecular Dynamics Model (IQMD)
is used to investigate the origin of the collective transverse velocity 
observed in heavy ion experiments. We find that there are three contributions to
this effect: initial-final state correlations, potential interactions and
collisions. For a given nuclear equation of state (eos) the increase of the 
transverse velocity with increasing beam energy is caused by the potential part.
For a given beam energy the collective transverse velocity is independent
of the nuclear eos but the relative contributions of potential and collisions
differ. In view of the importance of the potential interactions between
the nucleons it is not evident that the similarity of the radial velocities 
measured for fragments at beam energies below 1 AGeV and that for mesons at 
beam energies above 2 AGeV is more than accidental.
}\end{abstract}
\end{quote}}
\maketitle

A major goal of heavy ion collision studies  is to 
extend our knowledge
of the properties of dense and hot nuclear
 matter. In the past these studies were concentrated on the production of 
secondary particles like pions or kaons and on anisotropies in the
momentum distribution like the directed
in-plane flow (bounce-off) or the out of plane flow (squeeze-out) 
\cite{cathechism,joerg} of nuclear matter. For recent reviews of the
different collective flows we refer to \cite{rr,rrr} 

Recently also 
the isotropic radial flow has gained large interest because at AGS 
and at CERN energies the existing data 
at mid-rapidity of most of the mesons and baryons can be well described by the 
assumption that 
the system is thermalized and expands with a collective velocity 
\cite{sta1,sta2}. 
Later a similar observation has been made at energies between  150A MeV and
1A GeV (see eg. \cite{radflow1} - \cite{radflow4}). There the fragments, 
observed at mid-rapidity,  show an energy which 
increases linearly with the fragment mass. 
At these low energies most of the pions come from the decay of 
nuclear resonances and therefore cannot be used to determine a transverse 
velocity and kaons are too rare to be of use for a such an analysis.

The radial flow velocity $\beta_t$ determined with help of the fragment kinetic energies
up to 1A GeV 
smoothly joints that determined with help of mesons and baryons starting from
11A GeV. This raised
the question whether both transverse collective flows have a common origin and 
whether their functional
dependence on the beam energy allows for conclusions about a possible transition
between hadronic matter and a quark gluon plasma. 

A common origin of the transverse flow of fragments and mesons is not at all
evident. 
Nucleons which are bound asymptotically in fragments do usually not suffer collisions 
with a large
momentum transfer, otherwise they are not bound anymore. On the contrary, the
creation of mesons requires a large momentum transfer. 

At beam energies below 1AGeV the radial flow seems to be  
a good candidate for measuring the 
nuclear equation of state. Increasing compression leads to an increasing
pressure which is released isotropically. 
The directed transverse flow in the reaction plane, which in principle also
depends on the pressure, is
known to be sensitive on the potential gradient. 
The squeeze-out of high energy particles perpendicular to the reaction plane
allows for a view inside the reaction zone. However, this effect is connected
to the geometry and the time scales of the system.

At CERN and AGS energies a detailed theoretical investigation of the origin of 
the transverse velocity
is difficult \cite{urqmd} because many details of the expansion of the hadronic matter 
like the cross sections between baryonic and mesonic resonances are
unknown. Therefore calculations rely on a multitude of assumptions. 

At energies
around 1  A GeV the situation is much better. Here we have models which simulate
the complete reaction and we find agreement between the results of these models
\cite{cathechism,joerg,hart1,xxx,bau}
and experiment for many observables. It is the aim of this article to take 
advantage of this situation and to investigate the origin of the transverse 
expansion  of fragments in detail. Details of the Quantum molecular dynamics
models, which are used for our study, 
can be found in the refs.\cite{joerg,hart1} where also the different
equations of state, mentioned in this article, are explained. 
We used for this work the IQMD version of the model with a variance of the 
Gaussian wave function in coordinate space of $L=4.33\rm fm^2$ and (if not
otherwise stated) with static potentials because the compressional 
energy is difficult to determine with momentum dependent interactions. 
A small value of $L$ yields a higher number of fragments than a large
one \cite{hart1,george} and reproduces better the experimental multiplicity.
This analysis is strongly based on the behavior of fragments. Therefore
we have chosen this small L value. The choice of this  value of $L$ 
yield only  a slight enhancement of the asymptotic value of $\beta_t$ 
for the 1 GeV case, where most of the analysis is done. 
For lower energies there is a visible 
reduction of the asymptotic $\beta_t$ \cite{george}.


In the QMD simulations there are three different
processes which contribute to the transverse velocity:
\begin{itemize} 
\item The nucleons which finally form a fragment A may have already
initially a
finite transverse velocity 
$$<v_i(A)> = 
\left\langle{1\over A}\sum_{j=1}^A \vec v^j_i \vec e_r\right\rangle,$$ 
where $v_i^j$ equals $v^j_i(t=0)$
and the sum runs over all nucleons which are finally entrained in the fragment
of size A. $<>$ means averaging over all fragments of size A observed at
mid-rapidity ($y^{\rm CM}/y_{\rm Proj.}^{\rm CM} \le 0.3$).
This velocity may depend on the size of the fragment. 
This type of initial -final state correlations have
been investigated by Gossiaux et al. \cite{gossiaux}.
\item The potential between the nucleons may contribute to the 
transverse velocity $$v_F(A) = \left\langle {1\over mA}
\int dt \sum_{j=1}^A -\vec \nabla U^j(t) \vec e_r \right\rangle.$$ To 
investigate its influence we sum up all contributions from the nuclear 
force to the transverse velocity. 
\item The collisions between the baryons may increase the transverse
velocity as well $$v_C(A) = \left\langle{1\over A}
\sum_{j=1}^A \sum_{all\ coll}(\vec v_j' + \vec v_2' - \vec v_j - \vec v_2)
\vec e_r\right\rangle$$
To calculate this contribution we add the transverse velocity
change in all individual collisions  $(v_j, v_2) \to (v_j' v_2')$ 
of the nucleons $j$ of a fragment of the size $A$. 
\end{itemize}

Before we start with a detailed investigation of the origin of the transverse 
flow $\beta_t$
we compare our calculations with the experimental results to make sure that the
gross features of the  transverse flow are well reproduced in our simulations. 
As it was shown in \cite{george} in IQMD the transverse expansion of the nuclear
matter shows a linear velocity profile for central collisions at intermediate
energies. This linear profile is used as input to expansion models which assume
the expansion to be due to a thermal and a collective part. 
Assuming $E_t(A)=1/2\,mA\beta_t^2+ 3/2 T$ we disentangle these
two parts by fitting  the transverse velocities at mid-rapidity  
($y^{\rm CM}/y_{\rm Proj.}^{\rm CM} \le 0.3$)
using two parameters $\beta_t$ and $T$: 
\begin{equation} 
v_t(A) = \sqrt{\frac{2E_t(A)}{Am}}= \sqrt{\beta_t^2 +{3T\over Am}},
\end{equation} 
 where $\beta_t$ is the collective transverse velocity,
$T$ is the temperature, $m$ is the mass of the nucleon and $A$ the mass number
of 
the fragments.
$v_t(A)$ is the mean transverse velocity of fragments with the mass number
$A$. 
We made two fits in order to be comparable with the two sets
of experiments\cite{radflow1,radflow2} : one which includes all masses up to ten
and a
second where only masses smaller than five are included.
\begin{figure}[hbtp]
\psfig{figure=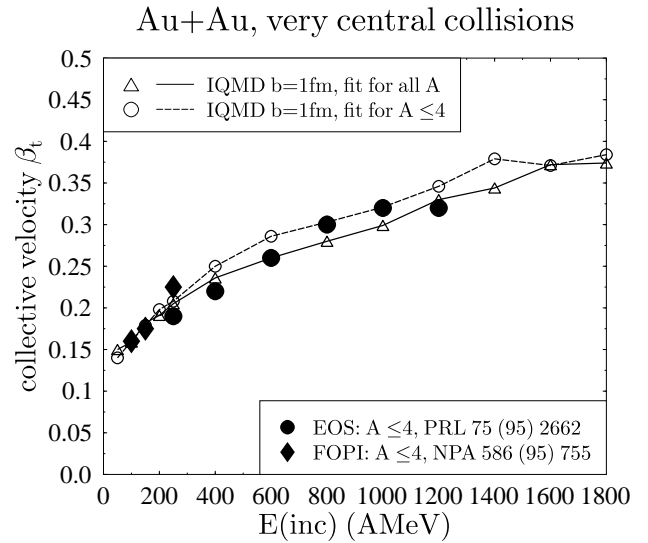,width=0.5\textwidth}
\caption{Transverse velocity $\beta_t$ obtained in IQMD simulations using a hard momentum
dependent interaction as compared to the different experiments.}
\end{figure}

Fig. 1 shows the results of the simulations for an impact parameter of b= 1 fm
using a hard equation of state with momentum dependent interactions. These
calculations are compared with the available experimental data.
The $\beta_t$ value of the different
experiments have been determined differently and in addition the impact
parameter cut is different. The difference between the data of\cite{radflow1} 
and of \cite{radflow2} is not understood yet. We observe that the simulations
follow very closely the experimental data. 


We see as well in fig.1 that even at 
the lowest beam energies a finite collective transverse velocity is observed. 
In this energy range Nebauer et al. \cite{neb} discussed on correlations
between the radial expansion and the directed transverse flow. However, they
studied semicentral collisions where the radial flow is small and the transverse
flow is strong. In this article we will only regard very central collisions
where the transverse flow is vanishing and the radial flow is strongest and 
analyze the influence of potential and collisions.

Fig. 2 shows the transverse velocities taken at mid-rapidity
for fragments produced in the reaction Au(250 AMeV)+Au at b=1fm.
\begin{figure}[hbtp]
\psfig{figure=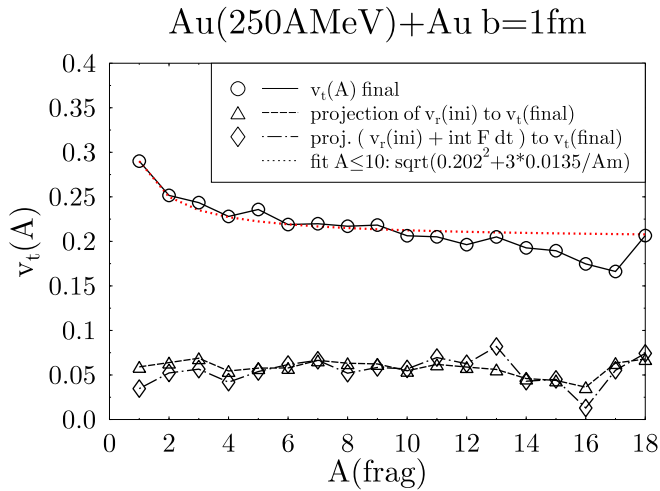,width=0.5\textwidth}
\caption{Transverse velocities $v_t(A)$ for the system Au(250 AMeV)+Au.
}
\end{figure}
The circles denote the final velocities of the fragments.
The dotted curve shows the result of a fit of the form 
of eq.1 and illustrates the method of determining the asymptotic
radial velocity as described above.

The triangles mark the initial total 
transverse velocity of the fragment nucleons. 
Clusters of nucleons which have a velocity
pointing away from the interaction zone have a higher chance to survive
the reaction without being destroyed than those clusters which have a momentum
which points into the reaction zone. For the diamonds we have added to the
initial transverse velocity the 
transverse velocity caused by the potential. 
The  influence of the potential on the transverse velocity is weak at that
energy. The difference between the diamonds 
and the circles, the total transverse velocity, is the transverse velocity caused by collisions. 
We see a continuous decrease of $v_t(A)$ with the fragment mass up to some
asymptotic value.  Thus, at 250
AMeV the transverse velocity has essentially two components: an initial
transverse velocity and that caused by collisions. The potential plays only a
minor role. 


It has now to be analyzed whether the vanishing influence of
the potential is a genuine effect or only valid for a beam energy of 250A MeV.
\begin{figure}[hbtp]
\psfig{figure=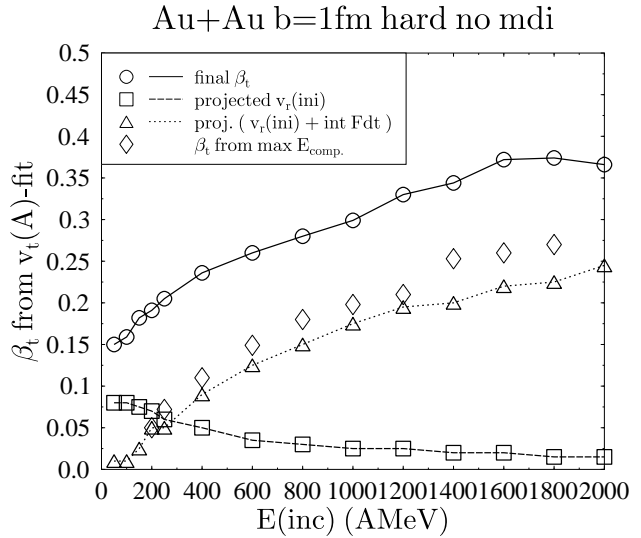,width=0.5\textwidth}
\caption{Energy dependence of the collective
transverse velocities $\beta_t$ for the system Au+Au and its different components.} 
\end{figure}
Figure 3 shows the asymptotic velocity and its decomposition into the different
contributions. 
We display  the initial transverse velocity of those nucleons which form
finally a fragment as well as the sum of the initial transverse velocity and that
caused by the potential. The difference between this curve and the asymptotic
value is the transverse velocity caused by collisions.   
For energies larger than 500A MeV we see an almost constant contribution from
the collisions ( of about $\beta = 0.1$) and from the initial velocity 
distribution of the fragments (of about $\beta = 0.04$). The
increase of $\beta_t$ is almost exclusively caused by the potential. Higher beam
energies yield a higher compression and therefore stronger forces. Only at
energies well above 2 GeV (where one may question the validity of our model)
we observe saturation. The low energy sector is more complicated. We see that
below 250A MeV the forces become attractive (the reaction develops towards a
scenario expected for deep inelastic collisions) and therefore they lower
the contribution from the initial velocity distribution. Collisions become
more and more Pauli blocked and do not contribute anymore. At high beam 
energies the  
nucleons which are finally entrained in fragments have been (in central collisions)
initially close to the surface at the back end of the nuclei, because there the number of
collisions they suffer is lowest \cite{gossiaux}. Also their initial 
momentum points away from the reaction zone. These
initial-final state correlations change at lower energies. There, due to the
increased Pauli blocking, projectile matter can traverse the target and vice
versa and hence the nucleons which are  finally found in fragments have initially a quite
uniform distribution over the nucleus. In \cite{neb} an initial final state
correlation is found for fragments emitted under $\theta_{CM}=90^\circ$.
\begin{figure}[btp]
\psfig{figure=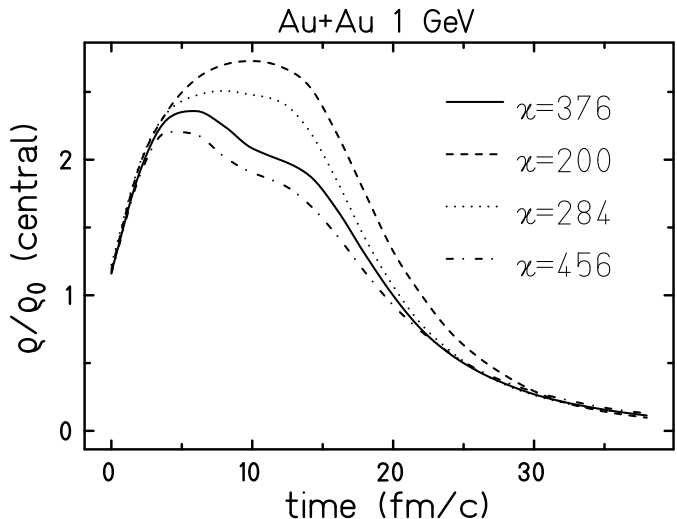,width=0.5\textwidth}
\caption{Time evolution of the density at the center of the reaction zone
for equations of state with different compressibilities.
}
\end{figure}

The physical origin of the energy
dependence of $\beta_t$ can also be studied by comparing it with the radial flow
expected if the total compressional energy is converted into a radial flow.
For this purpose we compare the energy difference 
$\Delta U = U_{max}- U_{ini}$ where $U_{max}$ is the total potential energy at the moment of the
highest compression whereas $U_{ini}$ is the initial potential energy of
projectile and target. The difference corresponds to the compressional energy 
needed to reach the state maximum compression.
 
$\Delta U$ is related to $\beta_t$ by  \begin{equation}
E=m+\Delta U= m\cdot \gamma= m \cdot \left( 1-\beta_t^2 \right)^{-1/2}
\end{equation}
The values are marked by diamonds. We see that its excitation function 
agrees to that of the contribution of initial $\beta_t$ plus 
potentials (triangles), which is the total radial flow minus the collisional
part. 
For higher energies this curve is parallel to that of the total radial flow
(circles) what verifies that the increase of the radial flow is caused
by the increasing compression, whereas the collisional contribution stays rather
constant.

All calculations discussed up to now have been performed using a hard
equation of state with momentum dependent interactions (mdi).
We now compare different equations of state at an energy of 1A GeV. 
\begin{figure}[hbtp]
\psfig{figure=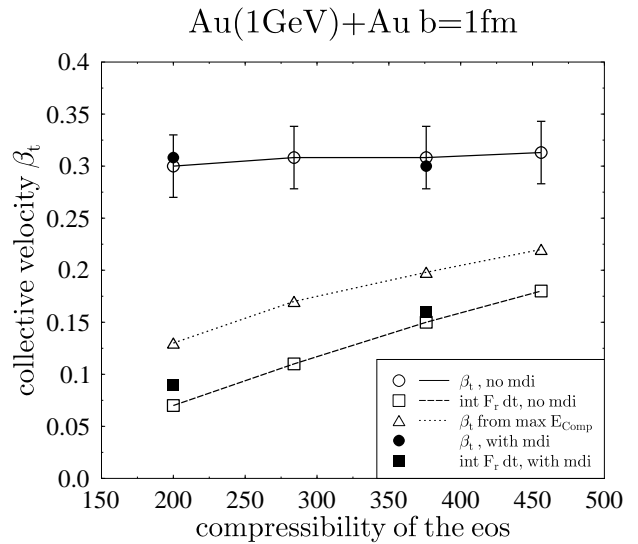,width=0.5\textwidth}
\caption{Dependence of the collective
transverse velocities $\beta_t$ on the compressibility
for the system Au(1A GeV)+Au and the
contribution of the potential part.}
\end{figure}
In the Skyrme ansatz the repulsion of the nuclear equation of state 
at high densities is connected to the nuclear compressibility.
Therefore we characterized in fig. 4 and 5 the equation of state by their
corresponding compressibility. Soft and hard eos correspond to $\kappa=200$ 
and $\kappa=376$ MeV, respectively.
Fig. 4 shows the time evolution of the density in the center of the 
reaction zone of a central Au+Au collision using different equations of state
without momentum dependent interaction.
We see that due to less repulsion the highest density is reached with 
the softest equation of state. Also the pressure is lower for this eos.
Consequently 
the system will stay for a longer time at maximum compression. A higher density causes a smaller mean free path and therefore an
enhanced number of collision. Both yield together a nearly linear dependence 
of the number of high energy collisions on the compressibility. 
Independent of the eos each high energetic nn collision contributes about the
same amount to the radial velocity. The softer the eos the more 
collisions contribute 
to the radial velocity and the higher is the kinetic pressure caused by 
the collisions. On the other hand
the repulsion of the potentials is weaker and causes a weaker pressure from
the potential part. Therefore it has to be checked which contribution to the
pressure is dominant and whether the dependence of the different contributions
to the total pressure counterbalance.

Figure 5 shows the asymptotic transverse velocities and the
contribution of the potential part as a function of the
compressibility for four different equations of state of 
Skyrme type. The difference is caused by collisions and
the initial state collective velocity. 
The latter contribution may be regarded as independent to the nuclear eos.
The values 
for a hard eos ($\chi $=376 MeV) resp. soft ($\chi $=200 MeV)
without momentum dependent interactions are very comparable to those with mdi. 
It is a remarkable result
that all equations of state yield about the same $\beta_t$
despite the contributions from the potential 
part are rather different. 
As already stated the potential part and the collisional part show different
dependences on the nuclear eos. Seemingly the differences counterbalance.

In fig.5 we show as well as open triangles the radial velocity one would obtain if the
compressional energy per nucleon is entirely converted into a radial velocity 
applying eq.2.: We see that this line
parallels that with the actual final transverse velocities caused by the
potential. Thus the increasing
compressional energy is responsible for the increase of the radial
velocity. The different absolute values are due to the fact that the system is
not in complete equilibrium and hence the compression is lowered due to the
not completely decelerated projectile and target matter. 

Using the IQMD model to investigate the collective transverse expansion 
velocity observed in heavy ion reactions we find, as in the experiments,
a continuous rise  with the beam energy up to energies of 1600A MeV. Whereas the
in-plane flow measures the potential gradient and the out of plane
squeeze the mean free path it was hoped that the radial flow is directly related
to the compressibility. This conjecture can only partially be substantiated.  
Indeed, the increase of the radial flow as a function of the bombarding energy 
is - for a given nuclear equation of state - due to the higher compression 
of the nuclear matter and not due to the rescattering of the fragment nucleons. 
The relative fraction of the potential and of the collisional contribution at a
given energy depends on the other hand on the compressibility of the equation 
of state. Even more, the sum of the potential and collisional contribution is
almost constant. Thus it is impossible to extract the compressibility from the
observation of the radial flow.

The collisional contribution to the radial flow should be rather similar 
for the mesons as well as for the baryons. At energies around 1 AGeV, however, 
the potential contributes in between 30\% and 60\% to the total radial velocity,
depending on the equation of state. This contribution is absent for the mesons
if created by string decay and not by baryonic interactions like $NN\rightarrow
\Delta N$.
Because no potentials are at hand to continue our investigations to higher
energies the doubts remain whether the one should expect a smooth transition
between the collective radial velocity observed 1 AGeV for fragments 
and that measured for mesons at higher (AGS and CERN) beam energies.  



\end{document}